\documentclass[twocolumn,showpacs,amsmath,amssymb,pre]{revtex4}

\usepackage{graphicx}   
\usepackage{dcolumn}    
\usepackage{bm}         
\newcommand{\rd}{{\rm d}}
\newcommand{\re}{{\rm e}}
\newcommand{\ri}{{\rm i}}

\begin{document}

\title[Periodic thermodynamics]
      {Energy flow in periodic thermodynamics}

\author{Matthias Langemeyer}
\author{Martin Holthaus}

\affiliation{Institut f\"ur Physik, Carl von Ossietzky Universit\"at,
	D-26111 Oldenburg, Germany}

\date{December 8, 2013}

\begin{abstract}
A key quantity characterizing a time-periodically forced quantum system
coupled to a heat bath is the energy flowing in the steady state through 
the system into the bath, where it is dissipated. We derive a general 
expression which allows one to compute this energy dissipation rate for 
a heat bath consisting of a large number of harmonic oscillators, and 
work out two analytically solvable model examples. In particular, we 
distinguish between genuine transitions effectuating a change of the
systems's Floquet state, and pseudo-transitions preserving that state;
the latter are shown to yield an important contribution to the total
dissipation rate. Our results suggest possible driving-mediated heating
and cooling schemes on the quantum level. They also indicate that a driven
system does not necessarily occupy only a single Floquet state when being 
in contact with a zero-temperature bath.   
\end{abstract}


 
\maketitle


\section{Introduction}

``Periodic thermodynamics'', a notion coined by W.~Kohn, refers to the
statistical physics of quantum systems which are driven by an arbitrarily 
strong time-periodic perturbation, and are weakly coupled to a heat
bath~\cite{Kohn01}. Such systems have been considered in the context of, 
among others, Rydberg atoms driven by a monochromatic coherent microwave field 
in the presence of noise~\cite{BlumelEtAl91}, open quantum systems under the 
influence of strong laser fields~\cite{BreuerPetruccione97}, driven dissipative 
quantum tunneling~\cite{GrifoniHanggi98}, and, more recently, degenerate Bose 
gases driven far from equilibrium~\cite{VorbergEtAl13}. In particular, Breuer 
{\em et al.\/} have emphasized the existence of a quasistationary distribution
of Floquet-state occupation probabilities to which the system relaxes in 
the long-time limit under the combined effect of the time-periodic force and 
the heat bath~\cite{BreuerEtAl00}; this line of investigation has been taken 
up by Ketzmerick and Wustmann with a detailed view on the classical-quantum 
correspondence~\cite{KetzmerickWustmann10}. Even in this steady state, energy 
is continuously being fed by the driving force into the system and transported 
to the bath, where it is dissipated. This steady-state energy flow is one of 
the most important quantities characterizing a time-periodically driven open 
quantum system. 
  
In the present paper we discuss the calculation of the energy dissipation
rate from a conceptual point of view. We employ a golden rule-type
perturbational approach which tends to gloss over certain theoretical 
details showing up in more elaborate treatments based on a Lindblad master
equation~\cite{GrahamHubner94,KohlerEtAl97,HoneEtAl09,OQS02,FlemingHu12}, 
but which yields the same results when the Born-Markov approximation is made, 
and which has the merit of making the physical content of the central 
expression~(\ref{eq:EDR}) for the energy dissipation rate particularly 
transparent. We proceed as follows: In Sec.~\ref{sec:2} we discuss the golden 
rule for transitions among Floquet states. Although this is already implicitly 
contained in previous works~\cite{BreuerEtAl00,HoneEtAl09} we here give a 
detailed derivation, since this approach is capable of some generalizations. 
We then use this golden rule in Sec.~\ref{sec:3} for deriving the energy 
dissipation rate for a time-periodically driven quantum system interacting 
with a thermal heat bath of harmonic oscillators. Here we distinguish between 
a contribution due to pseudo-transitions, which do not change the system's 
Floquet state, and the one due to genuine Floquet transitions. In the 
subsequent two sections we study two analytically solvable model systems, 
with emphasis placed on the connection between the dissipation rate and the 
ac Stark shift. In Sec.~\ref{sec:4} we briefly reconsider the linearly forced 
harmonic oscillator~\cite{BreuerEtAl00}, which shows a fairly uncommon feature:
All its quasienergy levels exhibit exactly the same ac Stark shift, so that 
the energy dissipation rate is entirely due to the pseudo-transitions. In 
contrast, the two-level system interacting with a circularly polarized
radiation field investigated in Sec.~\ref{sec:5} possesses a more generic
level response, and a correspondingly richer dissipation pattern. Some 
conclusions are drawn in the final Sec.~\ref{sec:6}.

\section{``Golden rule'' for transitions among Floquet states}
\label{sec:2} 

Consider a quantum system governed for times $t < 0$ by a Hamiltonian $H_0(t)$ 
which is periodic in time with period~$T$, 
\begin{equation}
	H_0(t) = H_0(t + T) \; ,
\label{eq:HAM}
\end{equation}
so that we have the Schr\"odinger equation
\begin{equation}
	\ri\hbar\frac{\partial}{\partial t} |\psi^{(0)}(t)\rangle
	= H_0(t) |\psi^{(0)}(t)\rangle
\label{eq:SGL}
\end{equation}
for $t < 0$. We assume this system to possess a complete set of 
square-integrable Floquet states, that is, particular wave functions of 
the form~\cite{Shirley65,Zeldovich67,Sambe73}   
\begin{equation}
	|\psi^{(0)}_n(t)\rangle = 
	|u_n(t)\rangle \exp(-\ri \varepsilon_n t/\hbar) \; ,	
\label{eq:FST}
\end{equation}
with real quasienergies $\varepsilon_n$ and Floquet functions $|u_n(t)\rangle$ 
which inherit the periodicity of their Hamiltonian,
\begin{equation}
	|u_n(t)\rangle = |u_n(t+T)\rangle \; .	
\end{equation}
The general solution to Eq.~(\ref{eq:SGL}) then takes the form of a 
superposition of these states,
\begin{equation}
	|\psi^{(0)}(t)\rangle = \sum_n c_n 
	| u_n(t)\rangle \exp(-\ri \varepsilon_n t/\hbar) \; ,
\label{eq:SUM}
\end{equation}
with coefficients $c_n$ which remain constant in time. This has the physically
appealing consequence that one can assign constant occupation probabilities
$|c_n|^2$ to the Floquet states, despite the Hamiltonian's explicit 
time-dependence. An additional perturbation acting on the system for $t > 0$ 
will then induce transitions among the Floquet states, so that one can ask how 
their occupation numbers change in response to that perturbation, {\em i.e.\/}, 
what the corresponding transition probabilities are; this is the question that 
will be tackled in the present section. 

The proposition concerning the existence of such Floquet states~(\ref{eq:FST})
involves some mathematical subtleties which narrow down the range of systems
to which the following deliberations can be applied rigorously, and which 
therefore deserve to be spelled out in some detail. These complications derive 
from a simple observation: Defining $\omega = 2\pi/T$, one has the obvious 
identity
\begin{eqnarray} & & 
	|u_n(t)\rangle \exp(-\ri \varepsilon_n t/\hbar) 
\nonumber \\	& = &
	| u_n(t)\rangle \re^{\ri r \omega t}
	\exp(-\ri [\varepsilon_n + r\hbar\omega] t/\hbar) \; , 
\label{eq:FAC}
\end{eqnarray}	   
where the function $|u_n(t)\rangle \re^{\ri r \omega t}$ again is periodic 
in time with period~$T$, if $r$ is any positive or negative integer. Thus, 
the separation of a Floquet state~(\ref{eq:FST}) into a periodic function 
$|u_n(t)\rangle$ and its Floquet multiplier $\exp(-\ri \varepsilon_n t/\hbar)$ 
is not unique: One is always free to choose any integer $r$, and then to 
replace $|u_n(t)\rangle$ by $|u_n(t)\rangle \re^{\ri r \omega t}$, if one 
simultaneously replaces $\varepsilon_n$ by $\varepsilon_n + r\hbar\omega$.
That is, in contrast to the energy of an eigenstate of some time-independent
Hamiltonian a quasienergy is not defined uniquely, but only up to an
integer multiple of $\hbar\omega$. For instance, one could factorize the
Floquet states such that all quasienergies fall into the ``first Brillouin
zone'' $-\hbar\omega/2 \le \varepsilon < +\hbar\omega/2$. It needs to be
stressed, however, that this procedure would merely be a matter of convention 
and other choices are equally possible: As long as one considers the full 
Floquet states~(\ref{eq:FST}), rather than Floquet functions and quasienergies
separately, the particular choice of the integer~$r$ involved in the formal
factorization~(\ref{eq:FAC}) is devoid of any significance. 

Nonetheless, this Brillouin-zone structure of the quasi\-energy spectrum is 
the root of severe mathematical difficulties. Namely, assume that the 
Hamiltonian has the natural form $H_0(t) = K + \lambda W(t)$, where $K$ defines
an ``unperturbed system'' on which a time-periodic influence $W(t) = W(t+T)$ 
acts with adjustable strength~$\lambda$. Then for $\lambda = 0$ the system's 
quasienergy spectrum consists of the energy eigenvalues of $K$, taken modulo 
$\hbar\omega$. Assuming further that $K$ possesses infinitely many discrete 
energy eigenvalues, the corresponding quasienergy spectrum of $H_0(t)$ for 
$\lambda = 0$ generically covers the entire energy axis densely. The decisive 
question then is whether the quasienergy spectrum still remains a dense pure 
point spectrum when $\lambda > 0$, or whether it becomes continuous: In the 
first case the sum expansion~(\ref{eq:SUM}) is to be taken literally, so that 
the system's wave function is strictly quasiperiodic in time, whereas in the 
second case the continuous quasienergy spectrum gives rise to diffusive energy 
growth~\cite{BunimovichEtAl91}. This question concerning the nature of the 
quasienergy spectrum of periodically time-dependent quantum systems is known 
as the ``quantum stability problem''~\cite{Combescure88,Howland92a}; since 
its solution heavily involves operator-theoretic versions of the
Kolmogorov-Arnold-Moser theorem, it has drawn substantial interest in 
mathematical physics. Along this line of research, an important rigorous 
result is due to Howland: Provided the gap between successive energy 
eigenvalues of $K$ grows sufficiently rapidly, and $W(t)$ is bounded, the 
quasienergy spectrum pertaining to $K + \lambda W(t)$ has no absolutely 
continuous component~\cite{Howland89}; this finding has later been generalized 
by Joye~\cite{Joye94}. Wishing to avoid unnecessary mathematical complications,
but still aiming at physically meaningful statements, we restrict ourselves to 
such systems which do not admit an absolutely continuous quasienergy spectrum. 
The remaining class of systems still includes interesting and important 
models such as linearly forced anharmonic oscillators with superquadratic 
potentials~\cite{Howland92b}, or, as a limiting case, the ``driven particle 
in a box''~\cite{BreuerEtAl00}.  

We now stipulate that the system~(\ref{eq:SGL}) be prepared in an individual 
Floquet state $n = i$ for $t < 0$, and then subjected to some perturbation
$V(t)$ with an arbitrary time-dependence for $t > 0$, so that the evolution
of the wave functions is given by   
\begin{equation}
	\ri\hbar\frac{\partial}{\partial t} |\psi(t)\rangle
	= \big( H_0(t) + V(t) \big) |\psi(t)\rangle
\label{eq:FSG}
\end{equation}
for $t > 0$. In order to set up time-dependent perturbation theory within 
the Floquet framework, we adapt the standard textbook reasoning for 
evaluating transitions between energy eigenstates~\cite{Baym69}: We introduce 
the time-evolution operator $U_0(t)$ of the unperturbed, periodically 
time-dependent system~(\ref{eq:SGL}), which obeys the equation
\begin{equation}
	\ri\hbar\dot{U}_0(t) = H_0(t) U_0(t) 
\end{equation}
with the initial condition $U_0(0) = 1$, and then employ this operator for 
transforming the wave functions $|\psi(t)\rangle$ to a Floquet-interaction 
picture by means of the relation 
\begin{equation}
	|\psi(t)\rangle = U_0(t) |\psi(t)\rangle_{\rm I} \; .	
\end{equation}
The transformed wave function $|\psi(t)\rangle_{\rm I}$ then evolves
according to the equation
\begin{equation}
	\ri\hbar\frac{\partial}{\partial t} |\psi(t)\rangle_{\rm I}
	= V_{\rm I}(t) |\psi(t)\rangle_{\rm I} \; ,
\end{equation}
where
\begin{equation}
	V_{\rm I}(t) = U_0^\dagger(t) V(t) U_0(t)	
\label{eq:UDF}
\end{equation}
denotes the perturbation operator transformed to the interaction picture, 
leading to the exact integral equation
\begin{equation}
	|\psi(t)\rangle_{\rm I} = |\psi(0)\rangle_{\rm I}
	+ \frac{1}{\ri\hbar} \int_0^t \! \rd \tau \, V_{\rm I}(\tau)
	|\psi(\tau)\rangle_{\rm I} \; .	
\label{eq:INT}	
\end{equation}
It may be helpful to point out the rationale underlying this procedure: One 
might likewise convert the original evolution equation~(\ref{eq:FSG}) into 
an integral equation, thus remaining fully within the Schr\"odinger picture; 
if one could solve that equation exactly the detour to the interaction picture 
were dispensable. However, the actual benefit of integral equations of the 
type~(\ref{eq:INT}) lies in the fact that they lend themselves to an iterative 
solution, leading to a Neumann series~\cite{Arfken85}. Then the interaction 
picture offers a tremendous advantage over the Schr\"odinger picture: Since the
norm of the integral operator on the right hand side of Eq.~(\ref{eq:INT}) is 
determined by the assumedly small perturbation $V_{\rm I}(t)$, the convergence 
properties of its formal series solution can be expected to be significantly 
better than those of the corresponding series obtained in the Schr\"odinger 
picture. Therefore, one may obtain acceptable approximations when terminating 
the interaction-picture Neumann series at low orders; in this sense the 
Floquet-interaction picture shares the virtues of the usual interaction 
picture employed in time-dependent perturbation theory~\cite{Baym69}. 
  
For computing the probability of a transition from the initial Floquet state 
$n = i$ to some final Floquet state $n = f$ we require the projections 
\begin{equation}
	\langle u_f(t) | \psi(t) \rangle = 
	\langle u_f(t) | U_0(t) | \psi(t) \rangle_{\rm I} \; .  
\end{equation}
At this point, the fact that the unperturbed Hamiltonian~(\ref{eq:HAM})
depends {\em periodically\/} on time becomes decisive: While in the 
case of an arbitrary time-dependence the evolution operator $U_0(t)$ would
have to be expressed as a time-ordered exponential~\cite{Baym69}, here we 
have the Floquet representation  
\begin{equation}
	U_0(t) = \sum_n \re^{-\ri\varepsilon_n t/\hbar}
	| u_n(t) \rangle \langle u_n(0) | \; ,
\label{eq:UFL}
\end{equation}
giving
\begin{equation}
	\langle u_f(t) | \psi(t) \rangle = \re^{-\ri\varepsilon_f t/\hbar} 
	\langle u_f(0) | \psi(t) \rangle_{\rm I} \; . 	
\end{equation}
To first order in $V$, the solution to the integral equation~(\ref{eq:INT})
now reads
\begin{equation}
	|\psi(t)\rangle_{\rm I} = |u_i(0)\rangle
	+ \frac{1}{\ri\hbar} \int_0^t \! \rd \tau \, V_{\rm I}(\tau)
	|u_i(0)\rangle \; .	
\label{eq:FOA}
\end{equation}
Inserting the definition~(\ref{eq:UDF}) and again utilizing the Floquet
representation~(\ref{eq:UFL}), one evaluates 
\begin{equation}
	\langle u_f(0) | V_{\rm I}(\tau) | u_i(0) \rangle =
	\re^{-\ri(\varepsilon_i - \varepsilon_f)\tau/\hbar}
	\langle u_f(\tau) | V(\tau) | u_i(\tau) \rangle \; ,
\end{equation}
from which we immediately obtain the desired transition probability $P_{fi}$
for $f \neq i$ to lowest order in the perturbation~$V$: 
\begin{eqnarray}
\label{eq:FTP}
	P_{fi} & = & 
	\left| \langle u_f (t) | \psi(t) \rangle \right|^2 
\\	& = &
	\frac{1}{\hbar^2} \left| \int_0^t \! \rd \tau \,
	\re^{\ri(\varepsilon_f - \varepsilon_i)\tau/\hbar}
	\langle u_f(\tau) | V(\tau) | u_i(\tau) \rangle \right|^2 \; .
\nonumber
\end{eqnarray}	 
The apparent similarity of this result to the corresponding expression for 
transition probabilities among energy eigenstates~\cite{Baym69} once again 
emphasizes the fact that in periodically time-dependent quantum systems the 
Floquet states take over the role which the energy eigenstates play in a 
system governed by a time-independent Hamiltonian.

To proceed, we assume that the perturbation is instantaneously switched on
at time $t = 0$ and then stays constant,  
\begin{equation}
	V(t) = \left\{ \begin{array}{cl}
		0 & ; \quad t < 0 	\\
		V & ; \quad t \ge 0   	\; . 	
			\end{array} \right. 
\end{equation}
Expanding the Floquet functions into Fourier series,
\begin{equation}
	| u_n(t) \rangle = \sum_{k = -\infty}^\infty
	| u_n^{(k)} \rangle \re^{\ri k \omega t} \; , 
\label{eq:FOC}
\end{equation}
we obtain the Floquet transition matrix elements in the form
\begin{eqnarray}
	\langle u_f(\tau) | V | u_i(\tau) \rangle & = &
	\sum_{k,j}\langle u_f^{(k)} | V | u_i^{(j)} \rangle 
	\re^{\ri(j-k)\omega t}
\nonumber \\	& = &	
	\sum_\ell \re^{\ri\ell\omega t} V_{fi}^{(\ell)} \; ,		
\end{eqnarray}
where
\begin{eqnarray}
	V_{fi}^{(\ell)} = \sum_{k} 
	\langle u_f^{(k)} | V | u_i^{(k+\ell)} \rangle \; , 
\label{eq:FME}
\end{eqnarray}
and Eq.~(\ref{eq:FTP}) yields the transition probabilities
\begin{equation}
	P_{fi} = 
	\frac{1}{\hbar^2} \left| \sum_\ell \int_0^t \! \rd \tau \,
	\re^{\ri(\varepsilon_f - \varepsilon_i + \ell\hbar\omega)\tau/\hbar}
	V_{fi}^{(\ell)} \right|^2 \; .
\end{equation}	 
When evaluating the squared sum, the cross-terms average to zero over a few 
cycles, so that one is left with
\begin{eqnarray}
	P_{fi} & \approx & \frac{t^2}{\hbar^2} \sum_\ell
	\frac{\sin^2\!\Big(
	(\varepsilon_f - \varepsilon_i + \ell\hbar\omega)t/2\hbar)\Big)}
	{\Big((\varepsilon_f - \varepsilon_i + \ell\hbar\omega)t/2\hbar\Big)^2}
	\left| V_{fi}^{(\ell)} \right|^2
\nonumber \\	& \sim &
	\frac{2\pi}{\hbar} t \sum_{\ell} \left| V_{fi}^{(\ell)} \right|^2
	\delta(\varepsilon_f - \varepsilon_i + \ell\hbar\omega)
\label{eq:FGR}	
\end{eqnarray}
for intervals~$t$ which are, on the one hand, sufficiently long to 
allow for the replacement of the above squared sinc functions by delta 
distributions, but remain sufficiently short to justify the first-order 
approximation~(\ref{eq:FOA}) on the other. 

This expression~(\ref{eq:FGR}) constitutes the desired analog of the
``golden rule'' for transitions among Floquet states. Because each Floquet
state brings its own set of Fourier components~(\ref{eq:FOC}) into the 
dynamics, a transition $i \to f$ does not merely correspond to a single 
spectral line, but rather to a series of lines equally spaced by $\hbar\omega$,
being weighted with the squared sum~(\ref{eq:FME}) of the components' matrix 
elements.

\section{Energy flow through driven quantum systems}
\label{sec:3}

To take the next step, we imagine that the system described by $H_0(t)$ is 
coupled to an environment conforming to a Hamiltonian $H_B$. With $H_0(t)$ 
acting on the system's Hilbert space ${\mathcal H}_S$, and $H_B$ acting on 
the space ${\mathcal H}_B$ pertaining to the environmental degrees of freedom, 
the total Hamiltonian
\begin{equation}
	H(t) = H_0(t) \otimes 1 + 1 \otimes H_{\rm B} + H_{\rm int}	 
\end{equation}
then is defined on the product space ${\mathcal H}_S \otimes {\mathcal H}_B$.
Here we take the coupling to be of the form
\begin{equation}
	H_{\rm int} = V \otimes W \; ,
\end{equation}
with $V$ carrying the dimension of an energy, so that $W$ is dimensionless.
The previous reasoning leading to Eq.~(\ref{eq:FGR}) can then easily be
adapted: Assuming $H_B$ to possess eigenstates $|\varphi_n\rangle$ with
energies $E_n$, we have the replacements 
\begin{eqnarray}
	| u_\alpha(t) \rangle & \rightarrow & 
	| u_\alpha(t) \rangle \otimes | \varphi_n \rangle
\nonumber \\
	\varepsilon_{\alpha} & \rightarrow &
	\varepsilon_{\alpha} + E_n \; ,	
\end{eqnarray}
thus obtaining the rates
\begin{equation}
	\Gamma_{fi}^{mn} = \frac{2\pi}{\hbar} \sum_\ell 
	\left| V_{fi}^{(\ell)} \right|^2 \left| W_{mn} \right|^2
	\delta(E_m - E_n + \hbar\omega_{fi}^{\ell})
\end{equation}
for individual system-environment transitions $(i,n) \to (f,m)$, where 
$W_{mn} = \langle \varphi_m | W | \varphi_n\rangle$, and
\begin{equation}
	\omega_{fi}^{\ell} = (\varepsilon_f - \varepsilon_i)/\hbar
	+ \ell\omega
\label{eq:OFI}
\end{equation}
denotes the frequencies associated with the system's Floquet transition
$i \to f$. Note that our approach neglects a second-order shift of the 
quasienergies which is induced by the interaction with the environment, 
but usually is quite small~\cite{BlumelEtAl91}. In the following we restrict 
ourselves to environments which can be described as a ``heat bath'' consisting 
of a very large number of thermally occupied harmonic oscillators~\cite{OQS02}. 
Accordingly,  
\begin{equation}
	W = \sum_{\widetilde{\omega}} \left( b_{\widetilde{\omega}} +
	b_{\widetilde{\omega}}^\dagger \right)
\end{equation}
is a sum over all bath annihilation operators $b_{\widetilde{\omega}}$ and 
their adjoint creation operators $b_{\widetilde{\omega}}^\dagger$. We then 
have to distinguish two cases: If $E_n - E_m = \hbar\widetilde{\omega} > 0$,
so that the system gains the energy $\hbar\widetilde{\omega}$ while the bath 
is de-excited and a bath phonon of frequency $\widetilde{\omega}$ is 
annihilated, one has
\begin{equation}
	W_{mn} = \sqrt{n(\widetilde{\omega})} \; ,
\end{equation}		
where $n(\widetilde{\omega})$ is the initial occupation number of the bath 
oscillator involved in the transition. If, on the other hand, 
$E_n - E_m = \hbar\widetilde{\omega} < 0$, so that the system loses energy
to the bath and a phonon of frequency $\widetilde{\omega}$ is created, one
obtains 
\begin{equation}
	W_{mn} = \sqrt{n(|\widetilde{\omega}|) + 1} \; .
\end{equation}		
Therefore, introducing the appropriate thermally averaged 
quantities~\cite{BreuerEtAl00}
\begin{equation}
	N(\widetilde{\omega}) = \left\{ \begin{array}{lcll}
	\langle n(\widetilde{\omega})\rangle & = &
	\displaystyle
	\frac{1}{\re^{\beta\hbar\widetilde{\omega}}-1}
	& \text{ when } \widetilde{\omega} > 0
\\
	\langle n(-\widetilde{\omega})\rangle + 1 & = &
	\displaystyle
	\frac{\re^{-\beta\hbar\widetilde{\omega}}}
	     {\re^{-\beta\hbar\widetilde{\omega}}-1}
	& \text{ when } \widetilde{\omega} < 0 \; ,	
	\end{array}\right. 	
\label{eq:DNO}
\end{equation}
where $\beta = 1/(k_{\rm B}T)$ specifies the temperature~$T$ of the oscillator 
bath, and invoking its spectral density $J(\widetilde\omega)$, the total rate
\begin{equation}
	\Gamma_{fi} = \int_{-\infty}^{+\infty} \!\!\! 
	\rd \widetilde{\omega} \, J(|\widetilde{\omega}|) \,  
	\frac{2\pi}{\hbar^2} \sum_\ell \left| V_{fi}^{(\ell)} \right|^2 
	N(\widetilde{\omega}) \, 
	\delta(-\widetilde{\omega} + \omega_{fi}^{\ell})
\end{equation}
of bath-induced Floquet transitions $i \to f$ can be written as a sum,
\begin{equation}
	\Gamma_{fi} =  \sum_{\ell} \Gamma_{fi}^{(\ell)} \; ,
\label{eq:TFR}
\end{equation}
with partial rates being given by 
\begin{equation}
	\Gamma_{fi}^{(\ell)} = \frac{2\pi}{\hbar^2}  
	\left| V_{fi}^{(\ell)} \right|^2
	N(\omega_{fi}^{\ell}) J(|\omega_{fi}^{\ell}|) \; .
\label{eq:PTR}
\end{equation}	 
At this point, let us briefly check the consequences of factorizing 
the Floquet states $|u_n(t)\rangle\exp(-\ri\varepsilon_n t/\hbar)$ in 
different manners: If one replaces, for instance, $|u_f(t)\rangle$ 
by $|u_f(t)\rangle\re^{\ri r \omega t}$, and $\varepsilon_f$ by 
$\varepsilon_f + r\hbar\omega$ with arbitrary integer~$r$, the matrix 
elements~(\ref{eq:FME}) are relabeled to read $V_{fi}^{(\ell-r)}$, 
while the transition frequencies~(\ref{eq:OFI}) are referred to as
$\omega_{fi}^{(\ell-r)}$, both retaining their numerical values. Thus, the 
net effect of this replacement is a relabeling of $\Gamma_{fi}^{(\ell)}$ 
to $\Gamma_{fi}^{(\ell-r)}$. Therefore, the choice of the representative of 
each Floquet function, that is, the choice of the respective integer~$r$, 
merely is a matter of convenience.

Asking now for the steady-state distribution $\{ p_n \}$ of Floquet-state 
occupation probabilities $p_n$ which establishes itself under the 
influence of the bath, one has to solve the Pauli-type master 
equation~\cite{BreuerEtAl00,KetzmerickWustmann10,VorbergEtAl13} 	
\begin{equation}
	\dot{p}_n = 0 
	= \sum_m \big( \Gamma_{nm} p_m - \Gamma_{mn} p_n \big)
\label{eq:GME}
\end{equation}		
into which only the total rates~(\ref{eq:TFR}) enter. However, when asking 
for the rate~$R$ of energy dissipated in this steady state, a more detailed 
view is required: As emphasized in the above derivation, a partial rate 
$\Gamma_{fi}^{(\ell)}$ belongs to a transition over the course of which the 
system acquires the energy $\hbar\omega_{fi}^{\ell}$ from the bath. Therefore, 
the desired dissipation rate is determined by the sum over all these partial 
rates, each weighted with the occupation probability $p_i$ of the respective 
inital state, and multiplied by the energy $-\hbar\omega_{fi}^{\ell}$ lost to 
the bath:
\begin{equation}
	R = - \sum_{mn\ell} \hbar\omega_{mn}^{\ell} \, 
	\Gamma_{mn}^{(\ell)} \, p_n \; .
\label{eq:EDR} 
\end{equation}
It will be of interest for the subsequent discussion to observe that this 
dissipation rate~(\ref{eq:EDR}) can naturally be decomposed into two parts:
One contribution arising from genuine transitions $n \to m$ between 
{\em different\/} Floquet states $n$ and $m$, and another one due to the 
``pseudo-transitions'' $n \to n$. The former contribution is written as
\begin{equation}
	R_{\rm trans} = - {\sum_{mn\ell}}^\prime 
	\hbar\omega_{mn}^{\ell} \, \Gamma_{mn}^{(\ell)} \, p_n \; ,
\label{eq:ETR}
\end{equation}
where the prime at the sum sign is meant to enforce the condition $m \ne n$, 
whereas the latter takes the form 
\begin{equation}
	R_{\rm pseudo} = - \hbar\omega {\sum_{n,\ell>0}} \ell \, 
	\left( \Gamma_{nn}^{(\ell)} - \Gamma_{nn}^{(-\ell)} \right) p_n ,
\label{eq:CPT}
\end{equation}
having used $\omega_{nn}^{\ell} = \ell\omega$ in accordance with 
Eq.~(\ref{eq:OFI}). Since the diagonal matrix element 
$\langle u_n(t) | V | u_n(t)\rangle$ is real, its Fourier component 
$V_{nn}^{(-\ell)}$ equals the complex conjugate of $V_{nn}^{(\ell)}$. 
Observing further that the definition~(\ref{eq:DNO}) implies
\begin{equation}
	N(\ell\omega) - N(-\ell\omega) = -1
\end{equation}
for $\ell > 0$, we find 
\begin{equation}
	R_{\rm pseudo} = + \hbar\omega \sum_{n,\ell > 0}
	\frac{2\pi}{\hbar^2} \ell \left| V_{nn}^{(\ell)} \right|^2 
	J(\ell\omega) \, p_n \; .
\label{eq:EPS}
\end{equation}
Thus, the pseudo-transitions yield a positive contribution to the dissipation
rate, and make sure that energy is dissipated even in those cases in which
the bath-induced genuine Floquet transitions do not figure.

\section{The linearly forced harmonic oscillator}
\label{sec:4}

An unusually simple, but still quite instructive model system is provided by a 
particle of mass~$M$ which is moving in a one-dimensional quadratic potential 
with oscillation frequency~$\omega_0$ while being subjected to a sinusoidal 
force with amplitude~$F$ and angular frequency~$\omega \neq \omega_0$, as 
described in the position representation by the Hamiltonian   
\begin{equation}
	H(x,t) = -\frac{\hbar^2}{2M}\frac{\partial^2}{\partial x^2}
	+ \frac{1}{2}M\omega_0^2 x^2
	+ F x \cos(\omega t) \; .
\label{eq:DHO}	
\end{equation}
The construction of its Floquet states follows a route laid out by 
Husimi~\cite{Husimi53,terHaar75}: With $T = 2\pi/\omega$, let $\xi(t)$ be 
the $T$-periodic solution to the classical equation of motion
\begin{equation}
	M\ddot{\xi} = -M\omega_0^2\xi - F\cos(\omega t) \; ,
\end{equation}	
namely,
\begin{equation}
	\xi(t) = \frac{F}{M(\omega^2 - \omega_0^2)} \cos(\omega t) \; .
\label{eq:TPT}
\end{equation}
Then the solutions to the time-dependent Schr\"odinger equation are
given by superpositions of the wave functions
\begin{eqnarray}
\label{eq:FSO}
	\psi_n(x,t) & = & \chi_n\big(x - \xi(t)\big) \, 
	\re^{-\ri E_n t/\hbar}
\\ & \times &	 
	\exp\!\Big(\frac{\ri}{\hbar} \Big[ 
	M\dot{\xi}(t)\big(x - \xi(t)\big) 
	+ \int_0^t \! \rd \tau \, L(\tau) \Big] \Big) \; ,   
\nonumber
\end{eqnarray}
where $\chi_n(x)$ is an oscillator eigenfunction with energy 
$E_n = \hbar\omega(n + 1/2)$, and
\begin{equation}
	L(t) = \frac{1}{2}M\dot{\xi}^2 - \frac{1}{2}M\omega_0^2\xi^2
	- F\xi\cos(\omega t)
\end{equation}
denotes the classical Lagrangian of the system, evaluated along the
$T$-periodic trajectory~(\ref{eq:TPT}). The Floquet functions $u_n(x,t)$ 
and their quasienergies $\varepsilon_n$ are then easily obtained by 
extracting the component increasing linearly with time from the phase of 
the solutions~(\ref{eq:FSO}), giving~\cite{BreuerHolthaus89}   
\begin{eqnarray}
	u_n(x,t) & = & \chi_n\big(x - \xi(t)\big)
	\exp\!\Big(\frac{\ri}{\hbar} \Big[  
	M\dot{\xi}(t)\big(x - \xi(t)\big) 
\nonumber \\ 	& + &	
	\int_0^t \! \rd \tau \, L(\tau) 
	- \frac{t}{T}\int_0^T \! \rd \tau \, L(\tau) \Big] \Big)   
\label{eq:OFF}
\end{eqnarray}
and
\begin{eqnarray}
	\varepsilon_n & = & E_n - \frac{1}{T}\int_0^T \! \rd \tau \, L(\tau)
\nonumber \\	& = &		
	\hbar\omega_0(n + 1/2) + \frac{F^2}{4M(\omega^2 - \omega_0^2)} \; .
\label{eq:QEO}
\end{eqnarray}
This latter result~(\ref{eq:QEO}) expresses a peculiarity of the harmonic 
oscillator: All its states respond in the same manner to the external force,   
that is, all its energy levels exhibit precisely the same ac Stark shift
proportional to the square of the driving amplitude.

Imposing now a dipole-type interaction of the form
\begin{equation}
	V = \gamma x \; ,
\end{equation}	
the fact that the Floquet functions~(\ref{eq:OFF}) essentially 
are harmonic-oscillator eigenfunctions following the classical 
trajectory~(\ref{eq:TPT}) without change of shape greatly facilitates
the calculation of the required matrix elements~\cite{BreuerEtAl00}: 
\begin{eqnarray}
	\langle u_m | x | u_n \rangle & = & 
	\langle u_m(t) | x - \xi(t) | u_n(t) \rangle + \delta_{mn} \xi(t) 
\nonumber \\	& = & 
	\sqrt{\frac{\hbar}{2M\omega_0}} \Big(
	\sqrt{n} \, \delta_{m,n-1} + \sqrt{n+1} \, \delta_{m,n+1} \Big)
\nonumber \\	& & \
	+ \delta_{mn} \frac{F}{M(\omega^2 - \omega_0^2)} \cos(\omega t) \; .		
\end{eqnarray}
This expression provides the matrix elements $V_{fi}^{(\ell)}$, and therefore
allows one to determine the partial rates~(\ref{eq:PTR}). On the one hand, 
the only nonzero rates associated with genuine Floquet transitions are
\begin{eqnarray}
	\Gamma_{n-1,n}^{(0)} = \Gamma_{n-1,n} & = &
	\frac{\pi\gamma^2 J(\omega_0)}{\hbar M\omega_0} \, 
	\frac{n\,\re^{\beta\hbar\omega_0}}{\re^{\beta\hbar\omega_0} - 1}
\nonumber \\
	\Gamma_{n+1,n}^{(0)} = \Gamma_{n+1,n} & = &
	\frac{\pi\gamma^2 J(\omega_0)}{\hbar M\omega_0} \, 
	\frac{n + 1}{\re^{\beta\hbar\omega_0} - 1} \; .	
\label{eq:OGT}
\end{eqnarray}
On the other, each pseudo-transition $n \to n$ is characterized by the
two partial rates
\begin{eqnarray}
	\Gamma_{nn}^{(1)} & = &
	\frac{\pi\gamma^2 F^2 J(\omega)}
	     {2\hbar^2 M^2 (\omega^2 - \omega_0^2)^2} \, 
	\frac{1}{\re^{\beta\hbar\omega} - 1}
\nonumber \\
	\Gamma_{nn}^{(-1)} & = &
	\frac{\pi\gamma^2 F^2 J(\omega)}
	     {2\hbar^2 M^2 (\omega^2 - \omega_0^2)^2} \, 
	\frac{\re^{\beta\hbar\omega}}{\re^{\beta\hbar\omega} - 1} \; .
\label{eq:PPR}
\end{eqnarray}
Now the master equation~(\ref{eq:GME}) determining the quasi\-stationary 
Floquet distribution $\{ p_n \}$ takes the form
\begin{eqnarray}
	0 = \dot{p}_n & = & 
	\left( \Gamma_{n,n-1} p_{n-1} - \Gamma_{n-1,n} p_n \right)
\nonumber \\ 	& + & 	
	\left( \Gamma_{n,n+1} p_{n+1} - \Gamma_{n+1,n} p_n \right) \; .
\label{eq:MEO}
\end{eqnarray}  
Adding the corresponding equation for $\dot{p}_{n+1}$ effectuates an 
enhancement of the label~$n$ in the second bracket by one. Iterating 
this procedure, one deduces that the two brackets in this 
equation~(\ref{eq:MEO}) have to vanish individually, giving
\begin{equation}
	\frac{p_n}{p_{n-1}} = \frac{\Gamma_{n,n-1}}{\Gamma_{n-1,n}}
	= \re^{-\beta\hbar\omega_0} \; .
\label{eq:DBL}
\end{equation}
Evidently, the uncommon feature that the total rates~(\ref{eq:TFR}) consist,
for this particular system~(\ref{eq:DHO}), of only one partial rate ensures 
detailed balance of the Floquet transitions~\cite{BreuerEtAl00}, so that the 
quasistationary Floquet occupation probabilities are given by a geometric 
Boltzmann distribution:
\begin{equation}
	p_n = p_0 \re^{-n\beta\hbar\omega_0} 
\end{equation}    
for $n > 0$, while 
\begin{equation}
	p_0 = 1 - \re^{-\beta\hbar\omega_0} \; .
\end{equation}
Moreover, the fact that all quasienergies~(\ref{eq:QEO}) differ from the
unperturbed oscillator energies by the same ac Stark shift allows one to
express this distribution as a Boltzmann distribution over these 
quasienergy levels~\cite{BreuerEtAl00}: 
\begin{equation}
	p_n = \frac{1}{Z} \re^{-\beta\varepsilon_n}
\label{eq:SSO}
\end{equation}	
with the partition function
\begin{equation} 
	Z = \sum_{n=0}^{\infty} \re^{-\beta\varepsilon_n} \; .
\end{equation}
Hence, the quasistationary Floquet distribution equals the canonical 
equilibrium distribution, regardless of the driving force.  

Turning now to the energy dissipation rate in this steady state, the 
contribution~(\ref{eq:ETR}) due to the genuine transitions here becomes
\begin{eqnarray}
	R_{\rm trans} & = & -\sum_n \left( 
	\hbar\omega_{n+1,n}^{(0)} \Gamma_{n+1,n}^{(0)} p_n +
	\hbar\omega_{n-1,n}^{(0)} \Gamma_{n-1,n}^{(0)} p_n \right)
\nonumber \\	& = &
	-\hbar\omega_0 \sum_n \left( 
	\Gamma_{n,n-1} p_{n-1} -\Gamma_{n-1,n} p_n \right) \; ,		
\end{eqnarray}
which, in view of the detailed-balance condition~(\ref{eq:DBL}), 
reduces to $R_{\rm trans} = 0$. But there still remains the 
contribution~(\ref{eq:CPT}) due to the pseudo-transitions:
\begin{equation}
	R_{\rm pseudo} = - \hbar\omega \sum_{n}  
	\left( \Gamma_{nn}^{(1)} - \Gamma_{nn}^{(-1)} \right) p_n \; .
\end{equation}
Since now Eqs.~(\ref{eq:PPR}) yield 
\begin{equation}
	\Gamma_{nn}^{(1)} - \Gamma_{nn}^{(-1)} =
	- \frac{\pi\gamma^2 F^2 J(\omega)}
	       {2\hbar^2 M^2(\omega^2 - \omega_0^2)^2} \; ,
\end{equation}
the sum over $n$ corresponds to the normalization condition
$\sum_n p_n = 1$, giving
\begin{equation}
	R = R_{\rm pseudo} = 
	\hbar\omega \frac{\pi\gamma^2 F^2 J(\omega)}
	                 {2\hbar^2 M^2 (\omega^2 - \omega_0^2)^2} \; .
\label{eq:OFL}
\end{equation}   
Thus, we have a fairly complete description of the energy flow through 
the driven harmonic oscillator~(\ref{eq:DHO}): While the steady Floquet
distribution~(\ref{eq:SSO}) equals the thermal Boltzmann distribution over 
the unperturbed energy eigenstates for all parameters of the driving force, 
there is a continuous flow of energy through the system into the bath, which 
acts as an energy sink. This energy flow~(\ref{eq:OFL}) is entirely due to 
pseudo-transitions which preserve the Floquet state, as expressed by the 
fact that the spectral density in Eq.~(\ref{eq:OFL}) is to be evaluated at 
the driving frequency $\omega$, not at the oscillator frequency~$\omega_0$.  
The dissipation rate does not depend on the temperature of the bath, but 
grows quadratically with the driving amplitude, and becomes singular when 
the driving frequency approaches the oscillator frequency.

\section{The two-level system in a circularly polarized field}
\label{sec:5}

A two-level system interacting with a circularly polarized monochromatic 
classical radiation field, as described by the Hamiltonian~\cite{Rabi37}
\begin{equation}
	H_0(t) = \frac{1}{2}\hbar\omega_0\sigma_z + \frac{\mu F}{2}
	\left( \sigma_x \cos\omega t + \sigma_y \sin\omega t \right) \; ,
\label{eq:TLS}
\end{equation}
defines a further analytically solvable model which, in spite of its quite
minimalistic appearance, is already able to reveal several generic features 
of the energy dissipation mechanism. Here $\sigma_x$, $\sigma_y$, and 
$\sigma_z$ denote the usual Pauli matrices~\cite{Baym69}, and $F$ quantifies 
the strength of the radiation field mode with frequency~$\omega$ which 
couples to the bare two-level system with a constant~$\mu$. Thus, the energy 
eigenvalues of the unperturbed system are $E_\pm = \pm \hbar\omega_0/2$, so 
that $\omega_0$ is the frequency of transitions between these bare levels. 
The formal simplicity of this model~(\ref{eq:TLS}) is deceptive; in fact, 
its dynamics are by far richer than those of the driven harmonic 
oscillator~(\ref{eq:DHO}).

\subsection{Floquet functions and quasienergies}

With the help of the Rabi frequency 
\begin{equation}
	\Omega = \sqrt{\delta^2 + (\mu F/\hbar)^2} \; ,
\label{eq:GRF}
\end{equation}	
where
\begin{equation}
	\delta = \omega_0 - \omega
\end{equation}
denotes the detuning of the driving frequency~$\omega$ from the bare 
transition frequency~$\omega_0$, the Floquet states of the driven 
two-level system~(\ref{eq:TLS}) take the form~\cite{HolthausJust94}  
\begin{equation}
	| \psi_{\pm}(t) \rangle = \re^{\mp\ri\Omega t/2}
	\frac{1}{\sqrt{2\Omega}} \left( \begin{array}{l}
        \pm\sqrt{\Omega \pm \delta} \; \re^{-\ri\omega t/2} \\
     	\phantom{\pm}\sqrt{\Omega \mp \delta} \; \re^{+\ri\omega t/2}
	\end{array} \right) \; .
\end{equation}
From these states we split off the Floquet functions 
\begin{equation}
	| u_{\pm}(t)\rangle = \frac{1}{\sqrt{2\Omega}} 
	\left( \begin{array}{l}
	\pm\sqrt{\Omega \pm \delta} \\
	\phantom{\pm}\sqrt{\Omega \mp \delta} \; \re^{+\ri\omega t}
	\end{array} \right) \; ,
\label{eq:TFF}
\end{equation}
implying that the corresponding quasienergies read
\begin{equation}
	\varepsilon_{\pm} = \frac{\hbar}{2}(\omega \pm \Omega) \; .
\label{eq:TLQ}
\end{equation}		
Now an important distinction has to be made: If $\delta > 0$, so that
the driving frequency is detuned to the red side of the bare transition,
these quasienergies reduce to  
\begin{eqnarray}
	\varepsilon_+ & \to & +\hbar\omega_0/2
\nonumber \\
	\varepsilon_- & \to & -\hbar\omega_0/2 + \hbar\omega		
\end{eqnarray}
in the limit $F \to 0$ of vanishing driving amplitude. In contrast,
when the radiation field is blue-detuned and $\delta < 0$, these limits
are given by
\begin{eqnarray}
	\varepsilon_+ & \to & -\hbar\omega_0/2 + \hbar\omega
\nonumber \\
	\varepsilon_- & \to & +\hbar\omega_0/2 \; . 		
\end{eqnarray}
Thus, the Floquet state labeled by ``$+$'' exhibits a quasienergy which
{\em increases\/} monotonically with increasing amplitude~$F$; this state 
is continuously connected to the {\em excited\/} state of the bare two-level 
system when $\delta > 0$, and to its {\em ground\/} state when $\delta < 0$;
vice versa for the Floquet state labeled by ``$-$''. Expressed differently, 
the ac Stark shift of the two levels changes {\em qualitatively\/} when 
$\delta$ changes its sign:  As indicated in Fig.~1, the two levels repel each 
other with increasing amplitude in the case of red detuning, whereas they 
approach each other and cross for blue detuning. This characteristic behavior 
also leaves its traces in the energy dissipation rate.

\begin{figure}[t]
\includegraphics[angle=-90,width=7.0cm]{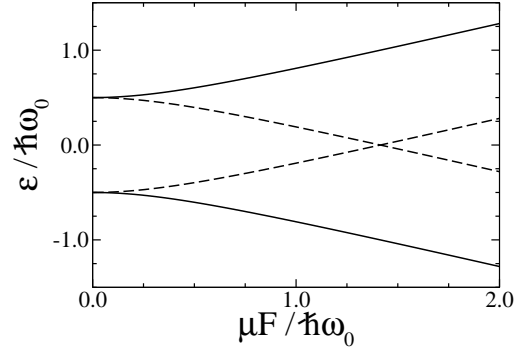}
\caption{Ac Stark shift for the two-level system driven by a circularly
	polarized radiation field: Shown 
	are those representatives of the quasienergies $\varepsilon$ which 
	connect continuously to the bare energy levels $\pm\hbar\omega_0/2$. 
	In the case of red detuning, when $\omega < \omega_0$, the two 
	quasienergies repel each other with increasing scaled amplitude 
	$\mu F/(\hbar\omega_0)$ (full lines: $\omega/\omega_0 = 0.5$), 
	whereas they approach each other and cross for blue detuning, when 
	$\omega > \omega_0$ (dashed lines: $\omega/\omega_0 = 1.5$).} 
\label{F_1}
\end{figure}

\subsection{Steady Floquet distribution}
\label{subsec:5b}

For modeling the system's coupling to the bath, we choose 
\begin{equation}
	V = \gamma \sigma_x \; ; 
\end{equation}
for simplicity, here we take the spectral density~$J$ of the bath to be 
constant. The Floquet functions~(\ref{eq:TFF}) readily yield the matrix 
elements
\begin{equation}	
	\langle u_+ | V | u_- \rangle = 
	\frac{\gamma}{2\Omega}\left( (\Omega + \delta) \re^{\ri\omega t}
	- (\Omega - \delta) \re^{-\ri\omega t} \right) \; ,
\end{equation}
possessing only two nonvanishing Fourier components
\begin{equation}
	V_{+-}^{(\pm 1)} = \pm \gamma \frac{\Omega \pm \delta}{2\Omega} \; ;
\end{equation}
likewise, one finds
\begin{equation}
	V_{-+}^{(\pm 1)} = \mp \gamma \frac{\Omega \mp \delta}{2\Omega} \; .
\end{equation}
For evaluating the partial rates~(\ref{eq:PTR}) we need to know the sign of
the associated transition frequencies $\omega_{fi}^\ell$, in order to resolve 
the distinction made in the definition~(\ref{eq:DNO}) of $N(\omega_{fi}^\ell)$.
Assuming $\Omega \le \omega$, Eq.~(\ref{eq:TLQ}) leads to  
\begin{eqnarray}
	\omega_{+-}^{1}  & = & \phantom{-} \Omega + \omega > 0
\nonumber \\
	\omega_{+-}^{-1} & = & \phantom{-} \Omega - \omega \le 0
\nonumber \\
	\omega_{-+}^{1}  & = & -\Omega + \omega \ge 0
\nonumber \\
	\omega_{-+}^{-1} & = & -\Omega - \omega < 0 \; ;	
\end{eqnarray}		
after introducing the convenient abbreviation
\begin{equation}
	\Gamma_0 = \frac{2\pi\gamma^2 J}{\hbar^2}
\end{equation}
one then finds the corresponding partial rates for the genuine
Floquet transitions, 
\begin{eqnarray}
	\Gamma_{+-}^{(1)} & = & \frac{(\Omega + \delta)^2}{4\Omega^2}
	\frac{\Gamma_0}{\re^{\beta\hbar(\omega + \Omega)} - 1}
\nonumber \\
	\Gamma_{+-}^{(-1)} & = & \frac{(\Omega - \delta)^2}{4\Omega^2}
	\frac{\Gamma_0 \re^{\beta\hbar(\omega - \Omega)}}
	     {\re^{\beta\hbar(\omega - \Omega)} - 1}
\nonumber \\
	\Gamma_{-+}^{(1)} & = & \frac{(\Omega - \delta)^2}{4\Omega^2}
	\frac{\Gamma_0}{\re^{\beta\hbar(\omega - \Omega)} - 1}
\nonumber \\
	\Gamma_{-+}^{(-1)} & = & \frac{(\Omega + \delta)^2}{4\Omega^2}
	\frac{\Gamma_0 \re^{\beta\hbar(\omega + \Omega)}}
	     {\re^{\beta\hbar(\omega + \Omega)} - 1} \; .
\label{eq:PRL}
\end{eqnarray}
However, when $\Omega > \omega$, the signs of $\omega_{+-}^{-1}$ and
$\omega_{-+}^{1}$ are reversed, resulting in 
\begin{eqnarray}
	\Gamma_{+-}^{(-1)} & = & \frac{(\Omega - \delta)^2}{4\Omega^2}
	\frac{\Gamma_0}{\re^{\beta\hbar(\Omega - \omega)} - 1}
\nonumber \\
	\Gamma_{-+}^{(1)} & = & \frac{(\Omega -\delta)^2}{4\Omega^2}
	\frac{\Gamma_0 \re^{\beta\hbar(\Omega - \omega)}}
	     {\re^{\beta\hbar(\Omega - \omega)} - 1} \; .
\label{eq:PRG}
\end{eqnarray}
This information suffices to determine the quasistationary Floquet distribution
$\{ p_+, p_- \}$ for both cases: Starting from the master equation
\begin{equation}
	0 = \Gamma_{+-} p_- - \Gamma_{-+} p_+
\end{equation}
and inserting $p_+ = 1 - p_-$, one has
\begin{eqnarray}
	p_- & = & \frac{\Gamma_{-+}}{\Gamma_{-+} + \Gamma_{+-}}
\nonumber \\	& = &
	\frac{\Gamma_{-+}^{(1)} + \Gamma_{-+}^{(-1)}}
	     {\Gamma_{-+}^{(1)} + \Gamma_{-+}^{(-1)} 
	    + \Gamma_{+-}^{(1)} + \Gamma_{+-}^{(-1)}} \; .
\label{eq:MPM}
\end{eqnarray}
After some elementary calculation, this leads for $\Omega \le \omega$ to
\begin{equation}
	p_- = \frac{1}{2} + \frac{\Omega\delta
	\left[\cosh(\beta\hbar\omega) - \cosh(\beta\hbar\Omega) \right]}
	{(\Omega^2 + \delta^2)\sinh(\beta\hbar\omega) 
	- 2\Omega\delta\sinh(\beta\hbar\Omega)} \; ,
\label{eq:PML}
\end{equation}
whereas for $\Omega > \omega$ we find
\begin{equation}
	p_- = \frac{1}{2} + \frac{\frac{1}{2}(\Omega^2 + \delta^2)
	\left[\cosh(\beta\hbar\Omega) - \cosh(\beta\hbar\omega) \right]}
	{(\Omega^2 + \delta^2)\sinh(\beta\hbar\Omega) 
	- 2\Omega\delta\sinh(\beta\hbar\omega)} \; .
\label{eq:PMG}
\end{equation}

\begin{figure}[t]
\includegraphics[angle=-90,width=7.0cm]{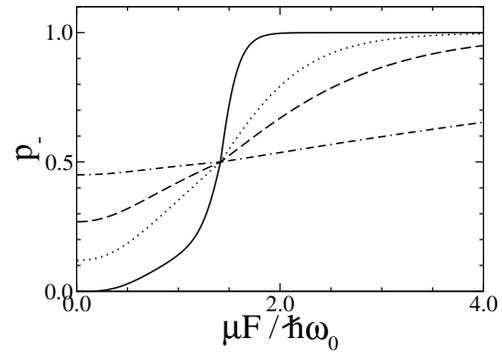}
\caption{Population of the Floquet state labeled ``$-$'' according to
	Eqs.~(\ref{eq:PML}) and (\ref{eq:PMG}) for $\omega/\omega_0 = 1.5$
	and scaled temperatures $k_{\rm B}T/(\hbar\omega_0) = 0.1$ (full
	line), $0.5$ (dotted), $1.0$ (dashed), and $5.0$ (dash-dash-dotted).} 
\label{F_2}
\end{figure}

Observe that $\Omega > \omega$ when   
$(\mu F/\hbar)^2 > 2\omega\omega_0 - \omega_0^2$, so that one always
ends up in the regime $\Omega > \omega$ when the scaled driving amplitude
$\mu F/(\hbar\omega_0)$ becomes sufficiently large. Eq.~(\ref{eq:PMG}) then 
implies $p_- \to 1$ for $\mu F/(\hbar\omega_0) \to \infty\;$: For any finite 
temperature of the bath, the Floquet state with a ``downward'' ac Stark shift 
will acquire all the population in the strong-forcing limit, regardless of 
whether this state is connected to the ground state or to the excited state 
of the bare two-level system in the opposite limit of vanishing driving 
amplitude. Figure~\ref{F_2} depicts $p_-$ vs.\ $\mu F/(\hbar\omega_0)$ for 
$\omega/\omega_0 = 1.5$, as corresponding to the ac Stark shift indicated
by the dashed lines in Fig.~\ref{F_1}. For low scaled temperature
$k_{\rm B}T/(\hbar\omega_0) = (\beta\hbar\omega_0)^{-1}$ the state labeled 
``$-$'', here being connected to the excited state of the bare system, 
naturally is almost unpopulated when $\mu F/(\hbar\omega_0) \ll 1$, 
but it accepts practically the entire population when 
$\mu F/(\hbar\omega_0) \gtrsim 2$. For higher temperatures the ``$-$''-state
carries more population already in the weak-driving regime, and the crossover
to the limit $p_- = 1$ is less pronounced, but this limit is approached with 
arbitrarily small deviation when $\mu F/(\hbar\omega_0)$ becomes large enough. 
If it were feasible to decouple the system from the heat bath at will, this 
would open up interesting heating and cooling schemes: Suppose than an 
ensemble of two-level systems is strongly driven, and that the ensemble's 
contact with the heat bath is disabled when  $p_- \approx 1$. If one then 
switches off the driving amplitude smoothly, the adiabatic principle for 
Floquet states~\cite{BreuerHolthaus89} guarantees that the value of $p_-$ 
remains practically unchanged. Therefore, if $\omega > \omega_0$ one obtains 
an ``ultrahot'' ensemble of bare two-level systems, with more or less all 
of its members being in the excited energy eigenstate at the end of the 
switch-off, whereas the final state would be an ``ultracold'' ensemble of 
two-level systems in their ground states when $\omega < \omega_0$.    

Another remarkable feature revealed by the steady-state occupation
probabilities~(\ref{eq:PML}) and (\ref{eq:PMG}) is their low-temperature limit 
in the presence of the drive: When $\Omega > \omega$, Eq.~(\ref{eq:PMG}) simply
yields $p_{-} \to 1$ for $\beta\hbar\omega_0 \to \infty\;$; in this case the 
``$-$''-state carries all the population for vanishing bath temperature. In
contrast, when $\Omega < \omega$ and Eq.~(\ref{eq:PML}) applies, one has
\begin{equation}
	p_{-} \to \frac{1}{2} + \frac{\Omega\delta}{\Omega^2 + \delta^2}
	\quad \text{ for } \beta\hbar\omega_0 \to \infty \; .
\label{eq:LTL}
\end{equation}  
If one now additionally takes the limit of vanishing driving amplitude,
this expression~(\ref{eq:LTL}) properly reduces to
\begin{equation}
	p_{-} \to \frac{1}{2} + \frac{1}{2} \, {\rm sign}(\delta)
	\quad \text{ for } \mu F/(\hbar\omega_0) \to 0 \; , 
\end{equation}
so that $p_{-} \to 1$ when $\delta > 0$ and the ``$-$''-state becomes the bare 
ground state, whereas $p_{-} \to 0$ when $\delta < 0$ and the ``$-$''-state
connects to the excited energy eigenstate of the bare two-level system. 
However, for finite nonzero driving strength matching the condition 
$\Omega < \omega$, none of the two Floquet states can accept all the 
population at zero temperature. 

The physics behind this finding becomes clear if one takes the limit of
vanishing temperature already at the level of the partial rates (\ref{eq:PRL})
and (\ref{eq:PRG}): For both $\Omega < \omega$ and $\Omega > \omega$
Eq.~(\ref{eq:PRL}) gives, for $\beta\hbar\omega_0 \to \infty$,
\begin{equation}
	\Gamma_{+-}^{(1)} \to 0
	\quad , \quad  
	\Gamma_{-+}^{(-1)} \to \frac{(\Omega + \delta)^2}{4\Omega^2}
	\Gamma_0 \; .		 
\end{equation}
In addition, when $\Omega > \omega$ one obtains from Eq.~(\ref{eq:PRG}) 
\begin{equation}
	\Gamma_{+-}^{(-1)} \to 0 
	\quad , \quad
	\Gamma_{-+}^{(1)} \to \frac{(\Omega - \delta)^2}{4\Omega^2}
	\Gamma_0 \; . 		 
\end{equation}
Hence, $\Gamma_{+-} = \Gamma_{+-}^{(1)} + \Gamma_{+-}^{(-1)}$ goes to zero in 
this case $\Omega > \omega$, while $\Gamma_{-+}$ remains nonzero. Therefore, 
at very low temperatures the driven system can still undergo transitions 
from ``$+$'' to ``$-$'', but not back from ``$-$'' to ``$+$'', so that 
eventually all the population piles up in the ``$-$''-state. In contrast, 
when $\Omega < \omega$ one deduces
\begin{equation}
	\Gamma_{+-}^{(-1)} \to \frac{(\Omega - \delta)^2}{4\Omega^2}
	\Gamma_0 		 
	\quad , \quad
	\Gamma_{-+}^{(1)} \to 0		 
\end{equation}
from Eq.~(\ref{eq:PRL}), so that now {\em both\/} 
$\Gamma_{+-} = \Gamma_{+-}^{(-1)}$ and $\Gamma_{-+} = \Gamma_{-+}^{(-1)}$ 
remain nonzero, and transitions in both directions remain enabled even at 
vanishing temperature; using Eq.~(\ref{eq:MPM}) one easily recovers 
Eq.~(\ref{eq:LTL}) from the above rates. The fact that even at vanishing 
bath temperature both ``upward'' and ``downward'' transitions can remain 
active is a distinctive feature of periodic thermodynamics.

\subsection{Energy dissipation rate}

Collecting all nonvanishing contributions, the energy dissipation 
rate~(\ref{eq:EDR}) for the circularly forced two-level system~(\ref{eq:TLS})
becomes
\begin{eqnarray}
	R & = & \hbar\omega 
	\left( \Gamma_{++}^{(-1)} - \Gamma_{++}^{(1)} \right)(1 - p_-)
\nonumber \\		& + & \hbar\omega
	\left( \Gamma_{--}^{(-1)} - \Gamma_{--}^{(1)} \right) p_-	
\nonumber \\	& - & \hbar(\Omega + \omega) \, \Gamma_{+-}^{(1)} \, p_-
	- \hbar(\Omega - \omega) \, \Gamma_{+-}^{(-1)} \, p_-
\nonumber \\  	& - & 
	\hbar(-\Omega + \omega) \, \Gamma_{-+}^{(1)} \, (1 - p_-)
\nonumber \\	& - & 
	\hbar(-\Omega - \omega) \, \Gamma_{-+}^{(-1)} \, (1 - p_-) \; ,
\end{eqnarray}		
where the first two terms on the right-hand side stem from the
pseudo-transitions. Computing the associated partial rates
\begin{eqnarray}
	\Gamma_{\pm\pm}^{(1)} & = & \frac{\Omega^2 - \delta^2}{4\Omega^2}
	\frac{\Gamma_0}{\re^{\beta\hbar\omega} - 1}
\nonumber \\
	\Gamma_{\pm\pm}^{(-1)} & = & \frac{\Omega^2 -\delta^2}{4\Omega^2}
	\frac{\Gamma_0 \re^{\beta\hbar\omega}}
	     {\re^{\beta\hbar\omega} - 1}
\end{eqnarray}
and observing
\begin{equation}
	\Gamma_{\pm\pm}^{(-1)} - \Gamma_{\pm\pm}^{(1)}
	= \Gamma_0 \frac{\Omega^2 - \delta^2}{4\Omega^2} \; ,
\end{equation}
we obtain
\begin{equation}
	R_{\rm pseudo} = \frac{\hbar\omega\Gamma_0}{4} 
	\left(\frac{\mu F}{\hbar\Omega}\right)^2 \; ,
\end{equation}
having used the definition~(\ref{eq:GRF}) of~$\Omega$. Next, the contribution
of the genuine Floquet transitions is rearranged to read
\begin{eqnarray}
	R_{\rm trans} & = &
	\hbar\Omega \left( \Gamma_{-+}^{(-1)} + \Gamma_{-+}^{(1)} \right) +
	\hbar\omega \left( \Gamma_{-+}^{(-1)} - \Gamma_{-+}^{(1)} \right) 
\nonumber \\	& - &
	\hbar\Omega \left( \Gamma_{-+}^{(-1)} + \Gamma_{-+}^{(1)} +
			   \Gamma_{+-}^{(1)} + \Gamma_{+-}^{(-1)} \right) p_-
\nonumber \\	& - &
	\hbar\omega \left( \Gamma_{-+}^{(-1)} - \Gamma_{-+}^{(1)} +
			   \Gamma_{+-}^{(1)} - \Gamma_{+-}^{(-1)} \right) p_-		
	\; .
\end{eqnarray}
Now the equation~(\ref{eq:MPM}) for $p_-$ effectuates the cancellation of 
the terms proportional to $\hbar\Omega$, and after some algebra one arrives at
\begin{equation}
	R_{\rm trans} = \frac{\hbar\omega\Gamma_0}{4}
	\frac{(\Omega^2 - \delta^2)^2}{\Delta^2\Omega^2} 
	\sinh(\beta\hbar\omega) 
\label{eq:RGT}
\end{equation}
with
\begin{equation}
	\Delta^2 = \left(\Omega^2 + \delta^2\right)\sinh(\beta\hbar\Omega_>) 
	- 2\Omega\delta\sinh(\beta\hbar\Omega_<) \; ,  
\end{equation}
where $\Omega_>$ ($\Omega_<$) is the larger (smaller) of the two frequencies
$\Omega$ and $\omega$; recall that this quantity $\Delta^2$ also appears in 
the denonimator of the steady-state occupation probabilities (\ref{eq:PML}) 
and (\ref{eq:PMG}). In contrast to $R_{\rm pseudo}$, this dissipation 
rate~(\ref{eq:RGT}) caused by the genuine Floquet transitions does depend on 
the temperature of the bath. Adding the two contributions, we finally obtain 
the total dissipation rate
\begin{equation}
	R = \frac{\hbar\omega\Gamma_0}{4} 
	\left(\frac{\mu F}{\hbar\Omega}\right)^2 
	\left[1 + 
	\left(\frac{\mu F}{\hbar\Delta}\right)^2 \sinh(\beta\hbar\omega) 
	\right] \; .
\end{equation}  
Considering finite nonzero temperatures, this total dissipation rate evidently 
approaches a finite value determined solely by the pseudo-transitions in the 
strong-forcing regime,
\begin{equation}
	R \to \frac{\hbar\omega\Gamma_0}{4}
	\quad \text{ for } \mu F/(\hbar\omega_0) \to \infty \; , 
\label{eq:SFL}
\end{equation}
whereas it vanishes in the high-frequency limit,
\begin{equation}
	R \to 0 
	\quad \text{ for } \omega/\omega_0 \to \infty \; .
\end{equation}

\begin{figure}[t]
\includegraphics[angle=-90,width=7.0cm]{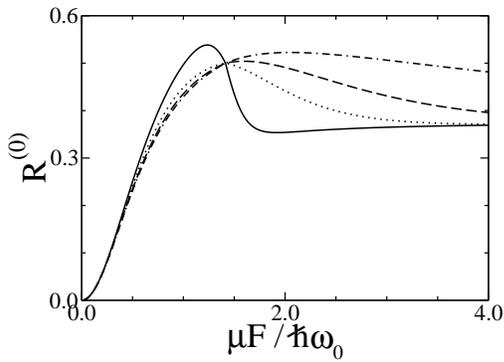}
\caption{Normalized energy dissipation rate 
	$R^{(0)} = R/(\hbar\omega_0\Gamma_0)$ for the situations considered
	in Fig.~\ref{F_2}, that is, for $\omega/\omega_0 = 1.5$ and scaled 
	temperatures $k_{\rm B}T/(\hbar\omega_0) = 0.1$ (full line), $0.5$ 
	(dotted), $1.0$ (dashed), and $5.0$ (dash-dash-dotted).} 
\label{F_3}
\end{figure}

\begin{figure}[t]
\includegraphics[angle=-90,width=7.0cm]{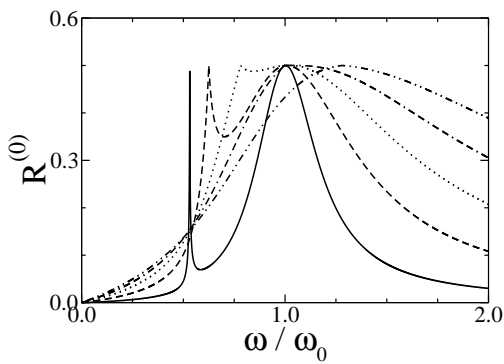}
\caption{Normalized energy dissipation rate 
	$R^{(0)} = R/(\hbar\omega_0\Gamma_0)$ for the scaled temperature 
	$k_{\rm B}T/(\hbar\omega_0) = 1.0$, and scaled driving srengths
	$\mu F/(\hbar\omega_0) = 0.25$ (full line), $0.5$ (dashed),
	$0.75$ (dotted), $1.0$ (dash-dash-dotted), and $1.25$ 
	(dash-dot-dotted).}
\label{F_4}
\end{figure}

In Fig.~\ref{F_3} we plot the normalized (dimensionless) rate 
$R^{(0)} = R/(\hbar\omega_0\Gamma_0)$ for precisely the cases studied
previously in Fig.~\ref{F_2}, that is, for $\omega/\omega_0 = 1.5$ and 
various scaled temperatures; here the approach to the strong-forcing 
limit~(\ref{eq:SFL}) is already recognizable for 
$\mu F/(\hbar\omega_0) \approx 4$. A better understanding of the 
non-monotonic behavior of these curves is obtained if one investigates the 
dependence of the dissipation rate on the driving frequency at fixed driving 
amplitude, as illustrated in Fig.~\ref{F_4}: For low driving amplitudes 
conforming to $\mu F/(\hbar\omega_0) < 1$ one has a broad resonance at about 
$\omega = \omega_0$ and a further, narrow resonance at $\omega = \Omega$,
which condition is equivalent to
\begin{equation}
	\frac{\omega}{\omega_0} = \frac{1}{2}
	\left( 1 + \left(\frac{\mu F}{\hbar\omega_0}\right)^2 \right) \; .
\end{equation}
Both resonances merge when $\mu F/(\hbar\omega_0) = 1$; for still higher 
driving amplitudes there is only one single, broad maximum of the dissipation
rate at a position determined mainly by the pseudo-transitions, located at 
$\omega/\omega_0 \approx \sqrt{1 + (\mu F/\hbar\omega_0)^2}$ in the
strong-forcing regime.

Once again, the limiting case of vanishing bath temperature merits special 
attention: When $\Omega > \omega$ one deduces
\begin{equation}
	R_{\rm trans} \to 0 
	\quad \text{ for } \beta\hbar\omega_0 \to \infty  
\end{equation}
from Eq.~(\ref{eq:RGT}), but when $\Omega < \omega$ we find
\begin{equation}
	R_{\rm trans} \to	\frac{\hbar\omega\Gamma_0}{4} 
	\left(\frac{\mu F}{\hbar\Omega}\right)^4
	\frac{1}{1 + (\delta/\Omega)^2}
	\quad \text{ for } \beta\hbar\omega_0 \to \infty \; .
\end{equation}
Thus, in the zero-temperature limit the genuine Floquet transitions do not 
figure when $\Omega > \omega$, but they do yield a finite contribution to the 
total dissipation rate when $\Omega < \omega$. This distinction evidently
matches the behavior of the steady-state occupation probabilities discussed 
at the end of the previous subsection.

\section{Conclusions}
\label{sec:6}

The concept of ``periodic thermodynamics''~\cite{Kohn01} implies that the 
steady state to which a quantum system relaxes in the presence of both a 
time-periodic driving force and a heat bath continuously delivers energy 
to the bath. The calculation of this steady-state energy dissipation rate, 
which we have outlined here for an harmonic-oscillator bath, reveals some 
peculiar features: Whereas the steady-state occupation probabilities are 
determined by a Pauli-type master equation into which only the rates for 
the genuine Floquet transitions enter~\cite{BreuerEtAl00}, the dissipation 
rate~(\ref{eq:EDR}) also incorporates the contribution~(\ref{eq:EPS}) from 
processes during which the bath energy changes by an integer multiple of 
$\hbar\omega$, while the system's Floquet state is left unchanged; these 
processes have been dubbed pseudo-transitions. Moreover, only the total 
rates~(\ref{eq:TFR}) embodying {\em all\/} Floquet transition frequencies 
show up in the master equation, whereas the evaluation of the dissipation 
rate~(\ref{eq:EDR}) requires the knowledge of the individual partial 
rates~(\ref{eq:PTR}). 

The example of the driven harmonic oscillator considered in Sec.~\ref{sec:4} is 
quite instructive insofar as it shows what does {\em not\/} happen in generic 
cases: For this particular model the total transition rates~(\ref{eq:OGT}) for 
the genuine transitions consist of only one partial rate, which implies that 
detailed balance still holds and the steady-state Floquet distribution equals 
the thermal Boltzmann distribution~\cite{BreuerEtAl00}; in addition, here 
the dissipation rate is given entirely by the pseudo-transitions. The study 
of the circularly forced two-level system performed in Sec.~\ref{sec:5} has 
demonstrated that the steady-state dynamics are substantially more involved 
when the bare energy levels of the driven system exhibit a nontrivial 
ac Stark shift. The features encountered here will also show up, multiply 
superimposed, in non-integrable systems which require numerical 
treatment~\cite{BreuerEtAl00,KetzmerickWustmann10,VorbergEtAl13}.
Particularly noteworthy is the fact that even in contact with a 
zero-temperature bath a time-periodically driven system does not necessarily 
occuply only one single Floquet state, as exemplified by the two-level model
in the regime $\Omega < \omega$.    
  
With a view towards future applications, the observations made at the end
of Subsec.~\ref{subsec:5b} might merit further investigations and
generalizations. The steady-state Floquet distribution which establishes
itself in the presence of the driving force may be preserved by decoupling
the bath and switching off the driving amplitude adiabatically; the resulting
state then can contain either more or even less energy than a stationary 
thermal state. Thus, as a matter of principle it seems possible to achieve
cooling by driving.

\begin{acknowledgments}
We thank Bettina Gertjerenken for helpful discussions. This work was supported 
by the Deutsche Forschungs\-gemeinschaft through Grant No.\ HO 1771/6-2.
\end{acknowledgments}

\end{document}